# Suppression of charge-ordering and appearance of magnetoresistance in a spin-cluster glass manganite $La_{0.3}Ca_{0.7}Mn_{0.8}Cr_{0.2}O_3$

*T. Sudyoadsuk[1], R. Suryanarayanan[2]\*, P. Winotai[1], L.E. Wenger[3]*

[1] Department of Chemistry, Mahidol University, Rama VI Road, Bangkok, 10400 Thailand

[2] Laboratoire de Physico-Chimie de l'Etat Solide, CNRS, UMR 8648, Bâtiment 414 Université Paris-Sud, 91405 Orsay, France

[3] Department of Physics & Astronomy, Wayne State University, Detroit, MI, 48201 USA

## ABSTRACT

The magnetic properties of electron-doped manganite $La_{0.3}Ca_{0.7}MnO_3$ and $La_{0.3}Ca_{0.7}Mn_{0.8}Cr_{0.2}O_3$ polycrystalline samples prepared by sol-gel technique have been investigated between 5 and 300 K in magnetic fields ranging from 0 to 5 T. The transition at 260 K, attributed to charge ordering in $La_{0.3}Ca_{0.7}MnO_3$, is completely suppressed in the Cr-substituted sample while the onset of a magnetic remanence followed by the appearance of a magnetic irreversibility at lower temperatures is observed in both samples. These features indicate that ferromagnetic clusters coexist with either an antiferromagnetic phase for $La_{0.3}Ca_{0.7}MnO_3$ or a spin-cluster glass phase for $La_{0.3}Ca_{0.7}Mn_{0.8}Cr_{0.2}O_3$ at the lowest temperatures. The exponential temperature dependence of the resistivity for the Cr-substituted sample is consistent with the small polaron hopping model for 120 K < $T$ < 300 K, while the data are better described by Mott's hopping mechanism for $T$ < 120 K. Whereas the parent compound $La_{0.3}Ca_{0.7}MnO_3$ is known to show no magnetoresistance, a large negative magnetoresistance is observed in the $La_{0.3}Ca_{0.7}Mn_{0.8}Cr_{0.2}O_3$ sample below 120 K. The appearance of the CMR is attributed to spin dependent hopping between spin clusters and/or between ferromagnetic domains.

# I. INTRODUCTION

The hole-doped manganites of $R_{1-x}D_xMnO_3$ (R = trivalent element = Bi, Y, La, Pr, Nd…; D = divalent elements = Pb, Ca, Sr, Ba) with $0 < x \leq 0.5$ have received renewed interest over the past decade due in part to the appearance of the so-called colossal magnetoresistance (CMR) effect.[1,2] This effect results from the application of a magnetic field driving a phase transition from a paramagnet insulator to a ferromagnetic metal in these manganites. The simultaneous presence of $Mn^{4+}$ and $Mn^{3+}$ ions in these compounds leads to the double exchange mechanism[3] for the $Mn^{4+}$-O-$Mn^{3+}$ coupling, which can account for the appearance of the ferromagnetism as well as the phase transition. However, several reports[4,5] have pointed out that the role played by the electron-phonon interaction mediated mainly by the Jahn-Teller distortion around the $Mn^{3+}$ environment should not be neglected. Others[6,7] have stressed the importance of phase separation in these compounds as numerous experimental observations indicate that these manganites are not magnetically homogeneous at the microscopic level. At the other end of the composition range for $x > 0.5$, the rare-earth and Bi-poor manganites exhibit a charge-ordering (CO) transition and become antiferromagnetic, charged-ordered insulators at low temperatures.[8,9] Since the formation of the CO state is accompanied by a lattice stiffening, the relationship between the electron-phonon coupling to the Jahn-Teller distortion and the magnetoresistance has been investigated in a variety of these electron-doped $Ln_{1-x}Ca_xMnO_3$ (Ln = lanthanides) compounds[10-17] with $0.5 \leq x \leq 1$. Although the appearance of a magnetoresistance (MR) effect depends on concentration of the lanthanide and hence the degree of stabilization of the CO state, the MR values are typically not as "colossal". For example, an MR value of about 70% is found for $La_{0.1}Ca_{0.9}MnO_3$[17] and 50% for $Bi_{0.125}Ca_{0.875}MnO_3$.[18] In another electron-doped manganite system, $Sm_{1-x}Ca_xMnO_3$, the substitution of another 3d-element into the Mn site[19-21] results in the disappearance of the CO and induces a CMR effect and ferromagnetic semimetallic ground state. Furthermore, Raveau et al.[22] have found that Cr and Co substitution on the charge-ordered insulator $Pr_{0.5}Ca_{0.5}MnO_3$ can induce a semiconductor-to-metal transition (SMT) in the absence of any applied magnetic field as well as a CMR. Among the 3d-elements, Cr substitution is particularly interesting as $Cr^{3+}$ is isoelectronic with $Mn^{4+}$ and is a non-Jahn-Teller ion. In addition, the nature of the magnetic interaction between $Cr^{3+}$-O-$Mn^{3+}$ is known to favor ferromagnetism through superexchange interaction. Hence, one might expect to induce a ferromagnetic phase in an otherwise antiferromagnetic, charge-ordered ground state for the electron-doped manganites depending upon the $Mn^{3+}/Mn^{4+}$ ratio. In fact, a ferromagnetic phase has been observed to coexist with the CO phase in Cr-substituted $Pr_{0.5}Ca_{0.5}MnO_3$ for low Cr concentrations at low temperatures.[22,23] This $Mn^{3+}/Mn^{4+}$ ratio as well as the orbital degree of freedom are thought to be important factors in determining whether a SMT can occur. Hence, it is interesting to investigate the structural, magnetic, and electrical transport properties of the electron-doped manganite $La_{0.3}Ca_{0.7}MnO_3$ with Cr substitution in which a ratio of $Mn^{3+}/Mn^{4+}$ smaller than both the hole-doped $La_{0.7}Ca_{0.3}MnO_3$ and the well-studied, charge-ordered $La_{0.5}Ca_{0.5}MnO_3$ compound with a $Mn^{3+}/Mn^{4+}$ ratio of 1. Magnetization measurements on a polycrystalline $La_{0.3}Ca_{0.7}Mn_{0.8}Cr_{0.2}O_3$ sample demonstrate that 20%Cr substitution in $La_{0.3}Ca_{0.7}MnO_3$ suppresses the charge-order transition observed at ~260 K and induces ferromagnetic spin-cluster glass behavior at lower temperatures. Though no SMT is observed with the application of a magnetic field, a large magnetoresistance is observed below 160 K due charge carriers hopping between these ferromagnetic clusters.

## II. EXPERIMENTAL DETAILS

Polycrystalline samples of nominal $La_{0.3}Ca_{0.7}MnO_3$ and $La_{0.3}Ca_{0.7}Mn_{0.8}Cr_{0.2}O_3$ composition were prepared by a sol-gel technique. Stoichiometric amounts of $La(NO_3)_3 \cdot 6H_2O$ (99%), $CaCO_3$ (99.5%), $Cr(NO_3)_3 \cdot 9H_2O$ (97%), and $Mn(CH_3COO)_2 \cdot 4H_2O$ (99%) were dissolved in a dilute $HNO_3$ solution with citric acid and ethylene glycol used as the chelating agents. The mixed solution was then heated until a dark-brown-colored resin material was formed. The resin was subsequently fired at 727 K and 1127 K in air to decompose the organic residual. The resultant powder was then ground, palletized, and sintered at 1527 K for 12 h.

Room-temperature powder X-ray diffraction (XRD) measurments on these samples were carried out with a Phillips PW 3020 diffractometer in the Bragg-Brentano geometry using $CuK_\alpha$ radiation. The XRD patterns recorded from $2\theta = 20°$ to $130°$ with a $0.02°$ step size and a counting time of 12 sec are shown in Figure 1 for both samples. These patterns are indexed to the $GdFeO_3$-perovskite structure with the space group *Pbnm* and do not show the presence of any other phases or precipitates of unreacted oxides. Further analysis by the Rietveld method using the *FULLPROF* program[24] resulted in lattice parameters of $a = 5.3761(2)$, $b = 5.3739(2)$, $c = 7.5715(1)$ Å ($V = 218.75(1)$ Å$^3$) for the $La_{0.3}Ca_{0.7}MnO_3$ sample and $a = 5.3558(3)$, $b = 5.3555(2)$, $c = 7.5703(3)$ Å ($V = 217.14(2)$ Å$^3$) for the $La_{0.3}Ca_{0.7}Mn_{0.8}Cr_{0.2}O_3$ sample. The decrease in the unit cell volume with the Cr substitution can be attributed to the presence of $Cr^{3+}$ since the ionic radius of $Cr^{+3}$ (0.61 nm) is smaller than that of $Mn^{3+}$ (0.65 nm). A summary of the Rietveld refinement data are listed in Table 1 and the *R*-factors seem to indicate a reliable fit to the $GdFeO_3$-perovskite structural model. The distortion of the $MnO_6$ octahedra in the Cr-doped sample ($\sigma^2_{JT}$ = 7.29×10$^{-6}$) is smaller than that of the parent sample ($\sigma^2_{JT}$ = 4.36×10$^{-5}$) where $\sigma^2_{JT}$ = $1/6$ $\Sigma\{[(Mn-O)_i-<Mn-O>]/<Mn-O>\}^2$. This would indicate that the $Cr^{3+}$ is residing at Mn sites rather than at the La sites.

The surface morphology and composition of the samples was examined by a scanning electron microscope (SEM) fitted with an energy dispersive spectrometer. The SEM micrographs showed a granular structure for both samples with a highly dense morphology with little or no porosity and typical grain sizes on the order of 1 μm. The EDS analysis indicated that the composition of the cations was quite close to the nominal one. Magnetization (*M*) measurements as a function of temperature (5 ≤ *T* ≤ 300 K) and field (0 ≤ *H* ≤ 5 T) were performed using a commercial Quantum Design SQUID magnetometer. The dc resistivity $\rho$ as a function of *T* and *H* was measured by a standard four-point probe method with silver paint used to make electrical contacts to the ceramic samples.

## III. RESULTS

### A. Magnetic properties

The temperature dependence of the magnetization *M/H* for the parent compound $La_{0.3}Ca_{0.7}MnO_3$ is presented in Fig. 2(a) for magnetic fields of 0.05 and 1.0 T. Three distinct features can be identified as being the onset of different magnetic behaviors: (i) the fairly sharp drop in the magnetization just below $T_{co}$ = 260 K, (ii) the appearance of a non-linear magnetic field dependence below $T_{fm}$ = 230 K, and (iii) the occurrence of an irreversibility between the zero-field-cooled (zfc) and field-cooled (fc) magnetization below $T_N$ = 170 K for the 0.1 T data. The sharp drop in the magnetization at 260 K is associated with the transition from a paramagnetic (PM) state to a charge-ordering (CO) state as this drop appears to be a magnetic characteristic of the CO transition in $La_{1-x}Ca_xMnO_3$ with *x* > 0.5. Also the temperature of 260 K for the CO transition is quantitatively similar to the value reported for $La_{0.333}Ca_{0.667}MnO_3$ from synchrotron x-ray and neutron powder diffraction results.[25] Previous phase diagrams[8,26] for the $La_{1-x}Ca_xMnO_3$ system indicate that the ground state for the *x* = 0.7 composition is antiferromagnetic at the lowest temperatures with the antiferromagnetic phase appearing around 170 K. Thus it is reasonable to identify the appearance of the irreversibility in the 0.1 Tesla measurements as the onset of antiferromagnetic ordering at the Néel temperature $T_N$. Although a modest field of 1.0 Tesla is sufficient to eliminate the irreversibility down to the lowest temperatures, there is a small, but noticeable, change in the slope of the higher field data (a small "hump") around 170 K which can be used to identify the antiferromagnetic ordering as well. The observation of a non-linear field dependence in the magnetization in the temperature region $T_N < T < T_{co}$ has not been previously reported for this composition. As will be shown for the Cr-substituted manganite sample, this non-linearity is probably due to a small magnetic remanence that arises from the presence of a ferromagnetic clusters forming at $T_{fm}$, which can co-exist with the antiferromagnetic state at lower temperatures. Such co-existence of small ferromagnetic clusters of Mn spins with an antiferromagnetic, charge-ordered structure has recently reported[27,28] in a similar composition range ($0.53 > x > 0.65$) for La$_{1-x}$Ca$_x$MnO$_3$.

The temperature dependence of the magnetization $M/H$ data for the La$_{0.3}$Ca$_{0.7}$Mn$_{0.8}$Cr$_{0.2}$O$_3$ compound is displayed in Fig. 2(b) for fields of 0.05, 0.5, 1.0 and 5.0 T. One immediately notes that the addition of Cr suppresses the sharp susceptibility peak associated with the charge ordering transition, yet retains the other two features of a non-linear $H$-field dependence at $T_{fm} = 136$ K and an irreversibility between the zfc and fc magnetization occurring at $T_{irr} \approx 80$ K for the lowest field. More definitive evidence for the complete suppression of the CO transition is found in the high-temperature plot of the inverse susceptibility as seen in Fig. 3. The magnetization follows a simple paramagnetic Curie-Weiss behavior above $T_{fm}$ with an effective magnetic moment $p_{eff}$ of 3.70 $\mu_B$. This value is about 7% less than the theoretical value of 3.98 $\mu_B$ assuming Cr$^{3+}$ completely substitutes for the Mn$^{3+}$ ions, an assumption in accordance with the conclusion reached by Rivadulla *et al.*[29] in the case of a hole-doped La$_{0.7}$Ca$_{0.3}$MnO$_3$ sample substituted with Cr. The positive paramagnetic Curie temperature of 65 K indicates that the predominate interaction between the Mn (Cr) spins must be ferromagnetic, which supports our previous speculation that the appearance of the non-linear magnetic field results from the formation of ferromagnetic clusters.

To verify that the magnetic behavior at $T_{fm}$ and below is associated with the presence of ferromagnetic clusters, $M$-vs-$H$ data between 80 K and 140 K was collected and are displayed in Fig. 4. While $M$ is a linear function of $H$ for $T = 140$ K and goes through the origin, a magnetic remanence and hysteresis are clearly observable for the $T \leq 120$ K data. However, the magnitude of the remanence is only ~ 0.01 $\mu_B$ /f.u., about 0.3% of the idealized saturated magnetic moment. This suggests that the ferromagnetic clusters involve only a small fraction of the Mn spins and correspondingly is rather short-ranged. This hypothesis is also consistent with the absence of any evidence for a spontaneous magnetization or a Curie temperature from Arrot plots. (Not shown.) One also notes from the *M-vs-H* plots that the magnetization can be described as $M = M_{rem} + aH$ for both the 100 K and 120 K data, suggesting the short-range ferromagnetic clusters co-exist with "paramagnetic-like" spins ($aH$) down to ~80 K.

At 80 K, the hysteretic behavior is different from the higher temperature data as the hysteresis is narrower with a smaller coercivity; and more importantly, the high-field behavior is very nonlinear. This non-linear behavior is even present down to 5 K as seen in Fig. 5. While this non-linear behavior is expected due to the presence of the irreversibility, it is also an indication of the onset for a different type of magnetic spin ordering at this temperature. One might speculate that antiferromagnetic ordering occurs in this Cr-substituted system, as only 10% of the "Mn" sites in $La_{0.3}Ca_{0.7}Mn_{0.8}Cr_{0.2}O_3$ are $Mn^{3+}$ since $Cr^{3+}$ is isoelectronic with $Mn^{4+}$. Thus, the probability (or number) of $Mn^{3+}$-$Mn^{4+}$ (or $Mn^{3+}$-$Cr^{3+}$) pairs coupled ferromagnetically by the double exchange (or superexchange) mechanism is very small as compared to a large number of antiferromagnetically coupled $Mn^{4+}$-$Mn^{4+}$ (or $Mn^{4+}$-$Cr^{3+}$) pairs that should exist. However, while modest magnetic fields are able to reduce the onset of the irreversibility to lower temperatures, the irreversibility is still noticeable below 20 K even for the 5 Tesla data as seen Fig. 2(b). The presence of the irreversibility for all fields and the lack of any clear slope change or small "hump" in any of the fc magnetization curves at 80 K signifies that the magnetic state for the Cr-doped $La_{0.3}Ca_{0.7}MnO_3$ is more similar to a ferromagnetic spin-cluster glass rather than an antiferromagnet. Additionally, the magnetic relaxation measured at 5 K is found to follow a stretched exponential functional form, which is qualitatively similar to the functional behavior typically associated with spin-glass systems. The time dependence of the magnetization $M(t)$ of the $La_{0.3}Ca_{0.7}Mn_{0.8}Cr_{0.2}O_3$ sample at 5 K after zero-field-cooling the sample from 300 to 5 K, applying a field of 5 Tesla, and then removing the field, is presented in Fig. 6. The fitting of the data to the stretched exponential function

$$M(t) = M(0) \exp[-(t/\tau)^{\beta}] \qquad (1)$$

results in a characteristic decay time $\tau$ of $9.92 \times 10^8$ sec and dispersion parameter $\beta$ of 0.36, which are quantitatively similar to those reported for a bilayer manganite.[30]  In addition to magnetic viscosity effects being a characteristic of spin-glass phenomena, spin-glass freezing temperatures defined as where the irreversibility occurs are known to shift to lower temperatures with increasing dc magnetic field in accordance with the present observation, and may be especially large for insulating spin glasses or cluster glasses.  More definitive evidence for the spin-glass freezing would be the observation of a shift of the spin-glass freezing temperature to higher temperatures with frequency in ac susceptibility measurements as observed for (Tb-La)$_{2/3}$Ca$_{1/3}$MnO$_3$[31] and Y$_{1-x}$Sr$_x$MnO$_3$ ($x$ = 0.5 and 0.6).[32]  Thus, in the absence of any detailed frequency-dependent ac susceptibility studies, one cannot totally exclude the possibility of some weak, non-collinear antiferromagnetic spin ordering with a weak viscosity behavior giving rise to the observed magnetic irreversibility in the La$_{0.3}$Ca$_{0.7}$Mn$_{0.8}$Cr$_{0.2}$O$_3$ sample.

## B.  Electrical transport properties

The electrical resistivity $\rho$ as a function of temperature ($H$ = 0 and 5 T) for the La$_{0.3}$Ca$_{0.7}$Mn$_{0.8}$Cr$_{0.2}$O$_3$ sample is displayed in Fig. 7.  In zero-field, the resistivity appears to exponentially increase from 0.23 $\Omega$-cm at room temperature to greater than $5 \times 10^7$ $\Omega$-cm below 20 K.  Although this behavior suggests that that sample remains essentially semiconducting over the entire temperature range, the resistivity data could not be fitted to a single exponential temperature dependence as seen in Fig. 8.  For $T > 120$ K, the small polaron hopping mechanism as discussed by Jaime *et al.*[33] to explain the resistivity for other manganites and described by $\rho(T) = A T \exp(E_{hop} / T)$ seems to provide the best fit to the data with an $E_{hop}$ value of ~100 meV as seen in the inset of Fig. 8.  This value for $E_{hop}$ is comparable to the value of 45 meV found for La$_{0.35}$Ca$_{0.65}$MnO$_3$ above its CO transition temperature.[26]  Below 120 K, the Mott's three-dimensional variable range hopping (3D-VRH) conduction process[34]

$$\rho(T) = \rho_o \exp[(T_o / T)^{1/4}] \qquad (2)$$

provides a better fit.  In Eq. (2), $\rho_o$ is a temperature independent parameter and $T_o$ ($= 5.24 \times 10^7$ K) is a measure of the degree of charge carrier localization.

Although no clear semiconductor-to-metal transition is observed in the resistivity data with the application of a 5 T field in Fig. 7, a "sizeable" negative magnetoresistance is discernible at low temperatures reaching a $\rho(5\ T)/\rho(0\ T)$ ratio of about 0.6 at 20 K. The temperature (20 K – 100 K) and field ( 0 – 5 T) dependences of the magnetoresistance $MR$ ( $\equiv \rho(H)/\rho(0)$) for the $La_{0.3}Ca_{0.7}Mn_{0.8}Cr_{0.2}O_3$ sample are more easily viewed in Fig. 9. While the resistivity is too large to be reliably measured below 20 K, the $MR$ is less than 1% above 160 K. In comparison, no magnetoresistance effect was observed in the parent compound $La_{0.3}Ca_{0.7}MnO_3$ in fields as high as 14 T.[35]

To further understand the origin of the $MR$ in this Cr-substituted manganite sample, fits to various field dependences and functional forms derived from different scattering models were attempted. Models based on a Born scattering[36] of charge carriers in a paramagnetic region where magnetic fluctuations are not weak or on a Kondo lattice[37] predict a low-field MR scaling of $\rho(H)/\rho(0) = 1 - C \cdot (M/M_s)^2$. Resulting plots of $\rho(H)/\rho(0)$-vs-$M^2$ exhibit some curvature with increasing magnetic field and $C$ values in the low magnetization limit ranging from 3 to 6 with decreasing temperatures. However, the coefficient $C$ in both of these models is predicted to be independent of temperature and field, which is not strictly observed for this sample. Instead, the most consistent fit over the entire temperature and field ranges results from a spin-dependent hopping model.[38] The magnetoresistance for this model is given by

$$\rho(H)/\rho(0) = exp[-W \cdot B_J^2(z) / k_B T ] \qquad (3)$$

where $z = g\mu_B JH/k_B T$ and $W$ is the energy barrier for hopping charge carriers between clusters. In the low magnetization limit, Eq. (3) becomes $\rho(H)/\rho(0) = 1 - C \cdot (M/M_s)^2$ where $C = (W/k_B T)$. As seen in Fig. 10, all of the magnetoresistance data falls on a single curve when plotted as a function of the variable $M^2/T$. Although there is some curvature in the data at the higher magnetization values, the slope from the low-magnetization and high-temperature region results in an energy barrier value $W$ equal to 170 meV assuming an $M_s$ value of 3.1 $\mu_B$/f.u. for the $La_{0.3}Ca_{0.7}Mn_{0.8}Cr_{0.2}O_3$ sample. This value is in very good agreement with the $W$ value of 150 meV found from the magnetoresistance results[38] of the double-layer manganite $(La_{0.4}Pr_{0.6})_{1.2}Sr_{1.8}Mn_2O_7$. Thus the magnetoresistance measured for $La_{0.3}Ca_{0.7}Mn_{0.8}Cr_{0.2}O_3$ resulting from charge carriers hopping between spin-aligned domains and between clusters with random mutual spin orientation is consistent with the ferromagnetic spin-cluster glass picture deduced from the magnetic properties.

## IV. DISCUSSION

As described in the preceding section, significant changes are observed in the magnetic and magnetotransport properties in $La_{0.3}Ca_{0.7}Mn_{0.8}Cr_{0.2}O_3$ from the parent $La_{0.3}Ca_{0.7}MnO_3$ compound. In particular, the compound $La_{0.3}Ca_{0.7}MnO_3$ exhibits a charge-order transition at $T \sim$ 260 K followed by an antiferromagnetic (AFM) transition at ~170 K. The charge-order transition observed at 260 K in the parent compound is completely suppressed by the Cr substitution such that the paramagnetic region extends down to 136 K. Moreover, our analysis of the zfc and fc magnetization data suggests that Cr-substitution induces a ferromagnetic spin-cluster glass behavior at the lowest temperatures rather than an antiferromagnetic ordering associated with the charged-ordered state found in $La_{0.3}Ca_{0.7}MnO_3$. The lack of any clear slope change or small "hump" in the fc magnetization at 80 K at the onset of the irreversibility and the qualitative stretched exponential form for the magnetic relaxation effect at 5 K are two features more consistent with a spin-cluster glass behavior than characteristic of antiferromagnetic ordering. Moreover, the appearance of a non-linear $M/H$ dependence in both samples at higher temperatures than the Néel or spin-glass freezing temperature suggests that ferromagnetic clusters or domains of small size are present as well. Further support for this speculation is seen in the $M$-$vs$-$H$ results for the $La_{0.3}Ca_{0.7}Mn_{0.8}Cr_{0.2}O_3$ sample as clear hysteretic behavior and magnetic remanence are observed. Also the paramagnetic Curie temperature being positive indicates that the dominant interaction between the moments at high temperatures is ferromagnetic. The ferromagnetic cluster formation may be somewhat surprising because of the limited number of ferromagnetically coupled $Mn^{3+}$-O-$Mn^{4+}$ (or $Mn^{3+}$-O-$Cr^{3+}$) configurations in $La_{0.3}Ca_{0.7}Mn_{0.8}Cr_{0.2}O_3$ and the even lower probability of this configuration extending over any sizeable length scale if the $Mn^{3+}$ sites are truly random. Nevertheless, the magnitude of the remanence being only ~ 0.01 $\mu_B$ /f.u., which is much less than the expected value of 3.1 $\mu_B$ /f.u. for a fully saturated magnetization, is in accordance with ferromagnetic clusters of relatively small size and number. Furthermore, other experimental evidence is rapidly being accumulated for magnetic and electronic texture on the mesoscopic length scale in the manganites. For example, the presence of such clusters or magnetic domains coexisting with the CO state has been observed[39] by Lorentz electron microscopy near the charge-ordering transition in $Nd_{0.5}Sr_{0.5}MnO_3$. Likewise, isolated magnetic domains[40] have been reported to appear in $La_{0.33}Pr_{0.34}Ca_{0.33}MnO_3$ thin films just before the SMT sets in. Therefore, while the magnetic properties for the parent $La_{0.3}Ca_{0.7}MnO_3$ compound are best described in terms of an antiferromagnetic, charge-order state co-existing with mesoscopic regions of ferromagnetic ordered Mn spins, the magnetic properties of the Cr-substituted manganite exhibit more of a spin-cluster glass behavior with small ferromagnetic domains being present. This interpretation is very similar to the conclusion drawn from the low-field magnetic properties of $Pr_{0.7}Ca_{0.3}MnO_3$, which closely resemble those of the $La_{0.3}Ca_{0.7}MnO_3$ compound. These authors[41] concluded that the magnetic properties arose from the presence of ferromagnetic clusters embedded within a non-ferromagnetic matrix.

The magnetotransport properties for the Cr-substituted $La_{0.3}Ca_{0.7}MnO_3$ sample are also consistent with a ferromagnetic spin-cluster glass behavior at low temperatures rather than with long-range spin ordering. Previous studies on manganites exhibiting spin-glass behavior display varying degrees of magnetoresistance ranging from a CMR[31] in $(La_{1/3} Tb_{2/3})_{2/3}Ca_{1/3}MnO_3$ as a result of a field-induced semiconductor-to-metal transition for fields greater than 5 T to no magnetoresistance effect[32] in $Y_{0.5}Sr_{0.5}MnO_3$ for fields as high as 4 T. Although the appearance of a large *MR* in the $La_{0.3}Ca_{0.7}Mn_{0.8}Cr_{0.2}O_3$ compound in itself does not indicate this system is more likely a spin-cluster glass with small ferromagnetic clusters, the $M^2/T$ dependence observed for the *MR*, which is attributable to a spin-dependent hopping between ferromagnetic clusters, is in accord with this magnetic structural picture.

## V. CONCLUSIONS

The effect of 20% Cr substitution at the Mn site on the magnetic and magnetotransport properties of the electron-doped manganite $La_{0.3}Ca_{0.7}MnO_3$ are reported. The transition at 260 K attributed to charge-ordering in $La_{0.3}Ca_{0.7}MnO_3$ is found to be suppressed in the Cr-substituted sample. Moreover, the magnetic properties of the $La_{0.3}Ca_{0.7}Mn_{0.8}Cr_{0.2}O_3$ compound resemble those of an inhomogenous spin-cluster glass with mesoscopic ferromagnetic clusters compared to the co-existence of an antiferromagnetic, charge-ordered phase with small ferromagnetic regions in $La_{0.3}Ca_{0.7}MnO_3$. The appearance of a large negative magnetoresistance in the Cr-substituted sample below 160 K is consistent with this structural picture as spin-dependent hopping between spin clusters or ferromagnetic domains appears to be the origin of the observed magnetoresistance behavior. And finally; we would like to point out that the data presented here would be of considerable significance in the light of recent work by Dagotto[42] wherein the author stresses the need to investigate the magnetic properties of *glassy state* and the associated CMR in manganites:

## Acknowledgments

TS, PW, and RS thank the Thai Royal Golden Jubilee program for financial support and to PERCH for partial financial support. TS and PW wish to thank A. Revcolevschi for the hospitality shown to them during their stay in Orsay.

**Table 1.** Crystallographic data for $La_{0.3}Ca_{0.7}O_3$ and $La_{0.3}Ca_{0.7}Mn_{0.8}Cr_{0.2}O_3$ from the Rietveld refinement of X-ray diffraction data. The space group *Pbnm* was used for all samples. La,Ca, and O1 occupy the $4c(x,y,1/4)$ site, Mn and Cr the $4a(1/2,0,0)$ site, and O2 the general $8d$ site. The numbers in parentheses are estimated standard deviations to the last significant digit, and $\sigma^2_{JT} = 1/6\ \Sigma\{[(Mn-O)_i-<Mn-O>\}/<Mn-O>]^2$.

| | | $La_{0.3}Ca_{0.7}MnO_3$ | $La_{0.3}Ca_{0.7}Mn_{0.8}Cr_{0.2}O_3$ |
|---|---|---|---|
| | $a$ (Å) | 5.3761(2) | 5.3558(3) |
| | $b$ (Å) | 5.3739(2) | 5.3555(2) |
| | $c$ (Å) | 7.5715(1) | 7.5703(3) |
| | $V$ (Å$^3$) | 218.75(1) | 217.14(2) |
| La/Ca | $x$ | -0.003(1) | -0.0021 |
| | $y$ | 0.020(2) | 0.020(1) |
| O1 | $x$ | 0.063(2) | 0.067(4) |
| | $y$ | 0.490(1) | 0.485(2) |
| O2 | $x$ | -0.278 (2) | -0.279(3) |
| | $y$ | 0.282(2) | 0.281(3) |
| | $z$ | 0.028(1) | 0.029(2) |
| | $B$ (Å$^2$) | 0.192(2) | 0.249(2) |
| | Mn-O1 $\times$ 2 (Å) | 1.923(2) | 1.928(4) |
| | Mn-O2 $\times$ 2 (Å) | 1.91(1) | 1.92(1) |
| | Mn-O2 $\times$ 2 (Å) | 1.94(1) | 1.92(1) |
| | Mn-O1-Mn | 159.58(9) | 157.88(2) |
| | Mn-O2-Mn | 161.3 (5) | 161.0 (6) |
| | $\sigma^2_{JT}$ | $4.36\times10^{-5}$ | $7.29\times10^{-6}$ |
| | $R_p$ | 14.0 | 16.1 |
| | $R_{wp}$ | 15.6 | 16.4 |
| | $R_B$ | 3.26 | 3.78 |
| | $\chi^2$ | 1.59 | 1.82 |

**FIGURE CAPTIONS**

**Fig.1** X-ray diffraction spectra of (a) $La_{0.3}Ca_{0.7}MnO_3$ and (b) $La_{0.3}Ca_{0.7}Mn_{0.8}Cr_{0.2}O_3$. Data points are indicated with solid circles, while the calculated patterns are shown as a continuous line. The positions of the reflections are indicated with the vertical lines below the patterns.

**Fig.2** The magnetic susceptibility *M/H* as a function of temperature *T* of (a) $La_{0.3}Ca_{0.7}MnO_3$ and (b) $La_{0.3}Ca_{0.7}Mn_{0.8}Cr_{0.2}O_3$. Both zero-field-cooled (zfc) and field-cooled (fc) data are shown for different magnetic fields *H*.

**Fig.3** Temperature dependence of the inverse susceptibility *H/M* of $La_{0.3}Ca_{0.7}Mn_{0.8}Cr_{0.2}O_3$. The solid line represents a Curie-Weiss fitting to the data.

**Fig.4** Field dependence of the magnetization of $La_{0.3}Ca_{0.7}Mn_{0.8}Cr_{0.2}O_3$ for $T \geq 80$ K.

**Fig.5** Field dependence of the magnetization of $La_{0.3}Ca_{0.7}Mn_{0.8}Cr_{0.2}O_3$ for $T \leq 100$ K.

**Fig.6** Time-dependence of the magnetization *M(t)* for $La_{0.3}Ca_{0.7}Mn_{0.8}Cr_{0.2}O_3$ at $T = 5$ K after magnetizing the sample in a field of 5 T and removing the field. The solid line is a fit to a stretched exponential function.

**Fig.7**    Resistivity $\rho$ of $La_{0.3}Ca_{0.7}Mn_{0.8}Cr_{0.2}O_3$ as a function of temperature in zero and 5 T magnetic fields.

**Fig.8**    Zero-field resistivity of $La_{0.3}Ca_{0.7}Mn_{0.8}Cr_{0.2}O_3$ as a function of $(1/T)^{1/4}$. The inset displays $\ln(\rho/T)$ *vs.* $T^{-1}$.

**Fig.9**    Normalized magnetoresistivity $\rho(H)/\rho(0)$ for $La_{0.3}Ca_{0.7}Mn_{0.8}Cr_{0.2}O_3$ at various temperatures as a function of magnetic field $H$.

**Fig.10**  Normalized magnetoresistivity $\rho(M)/\rho(0)$ for $La_{0.3}Ca_{0.7}Mn_{0.8}Cr_{0.2}O_3$ as a function of $M^2/T$.

Fig 1

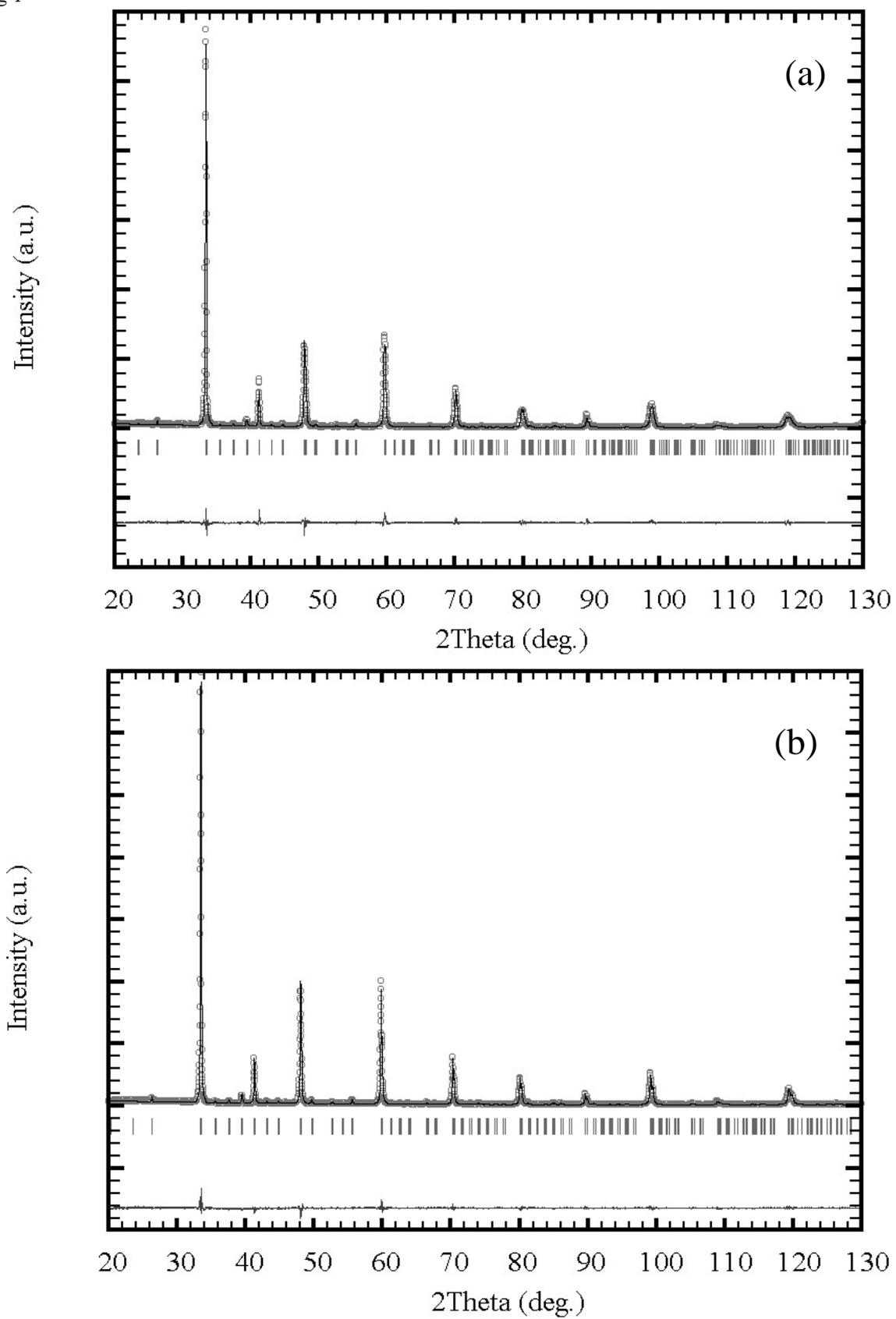

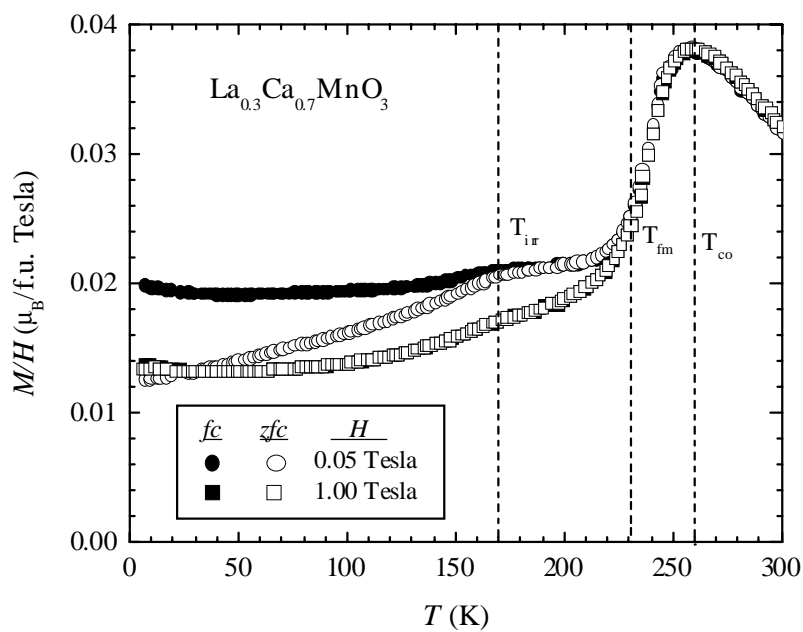

**Fig. 2(a)**

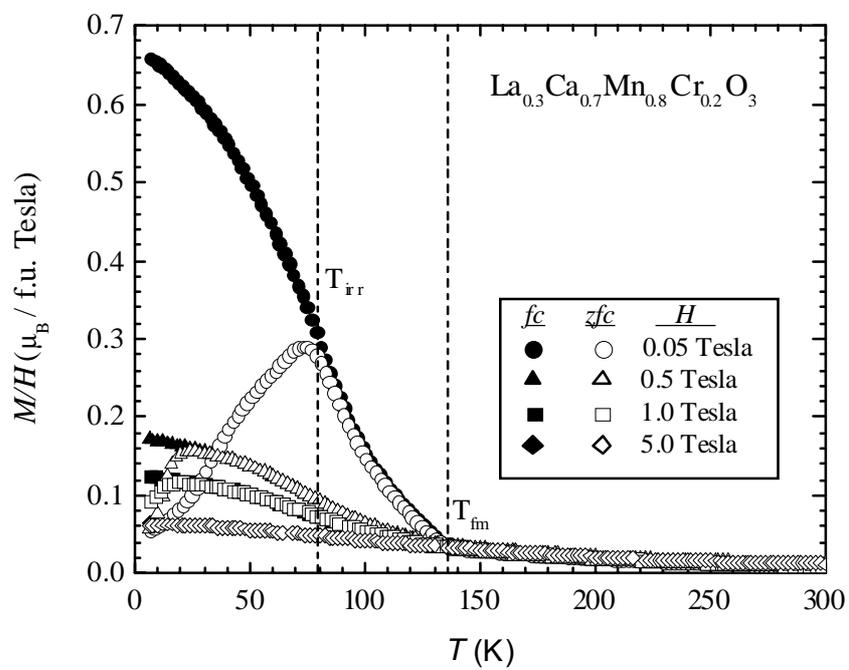

**Fig. 2(b)**

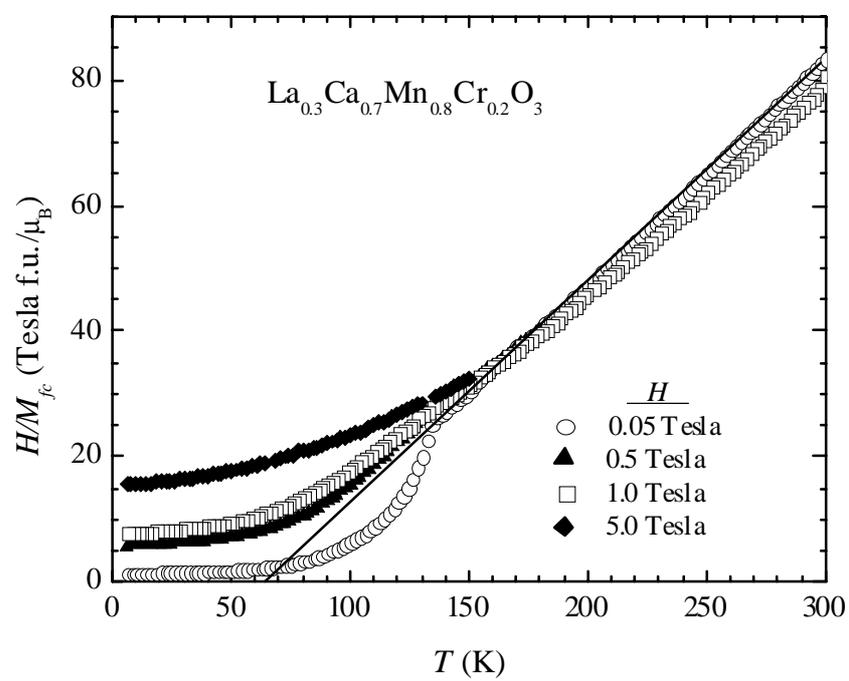

Fig. 3

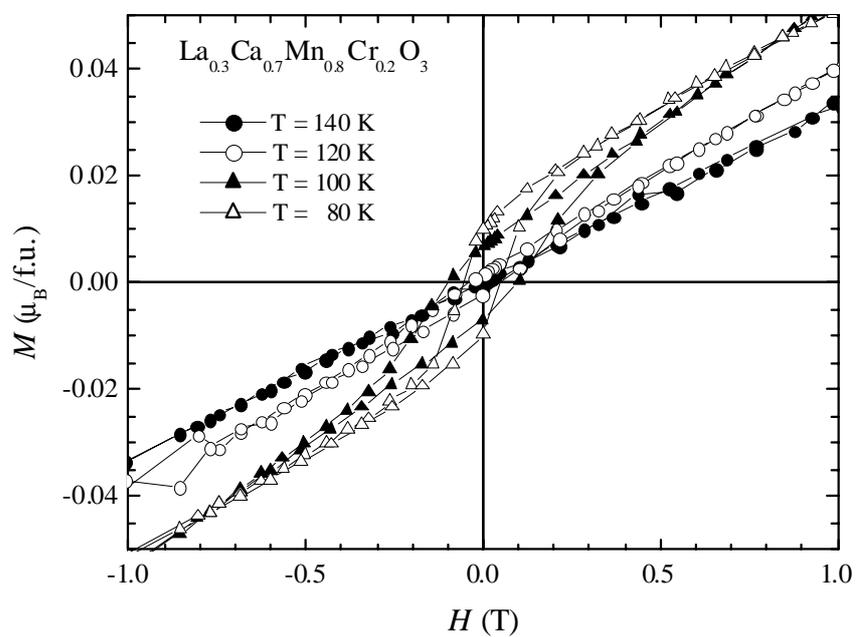

**Fig. 4**

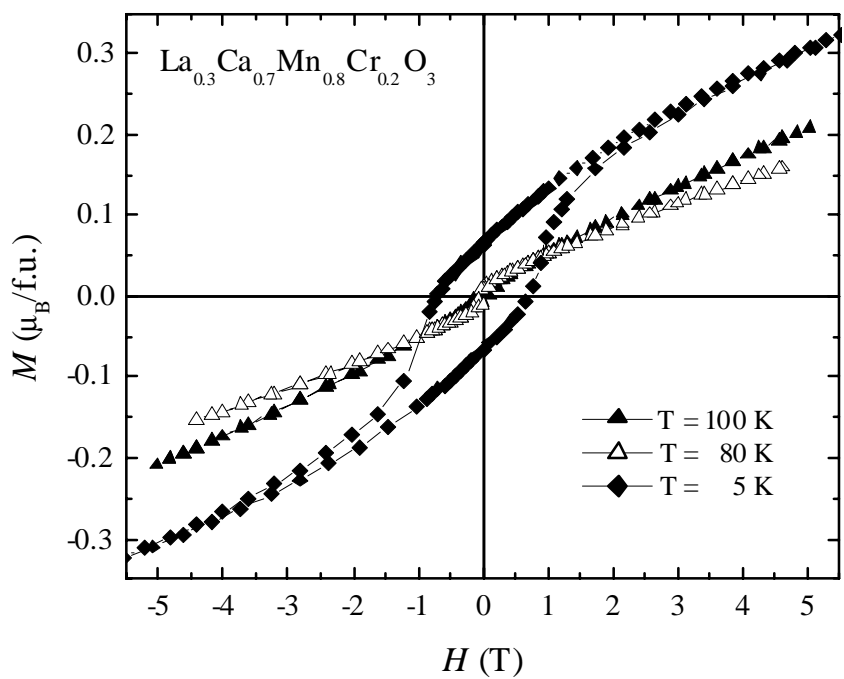

**Fig. 5**

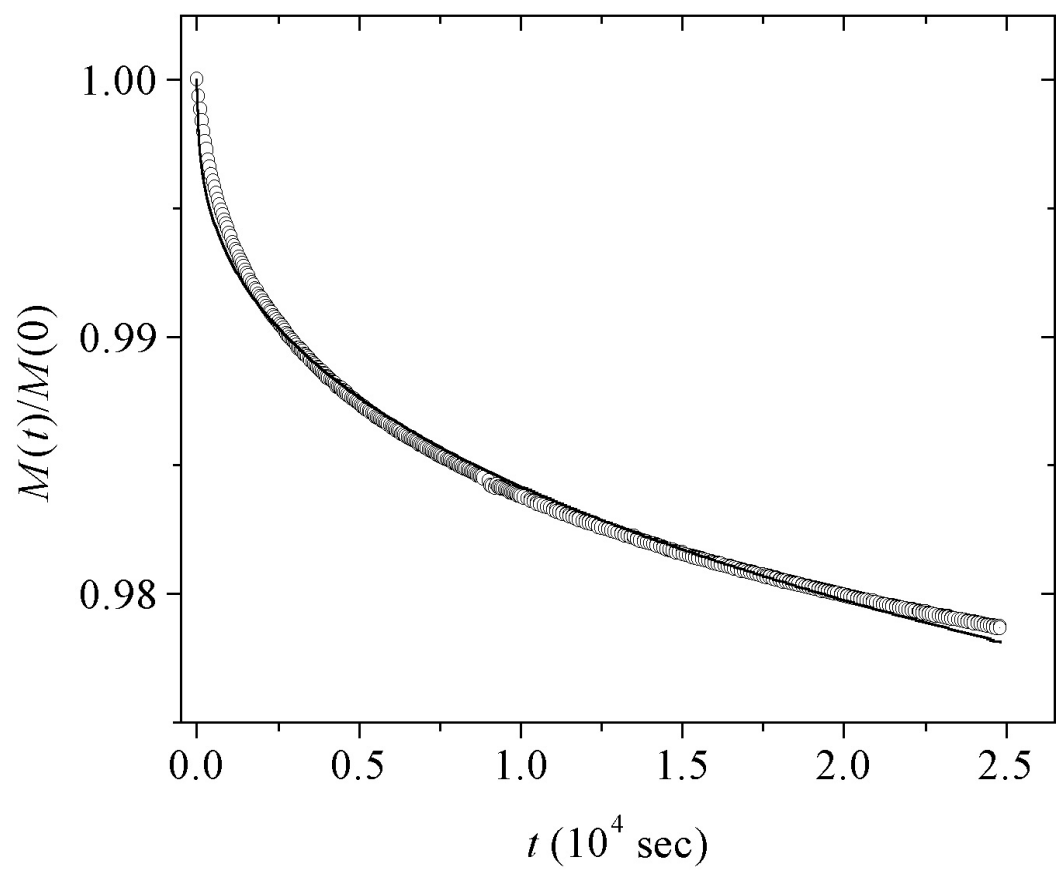

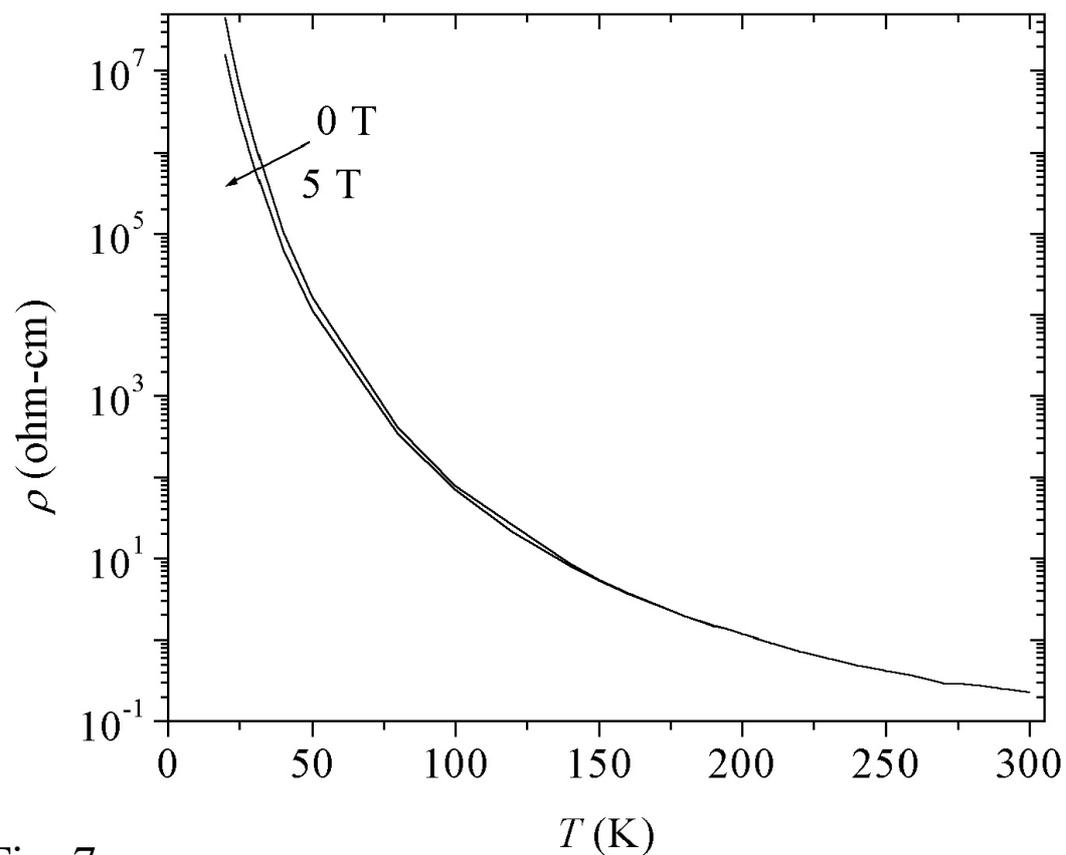

Fig. 7

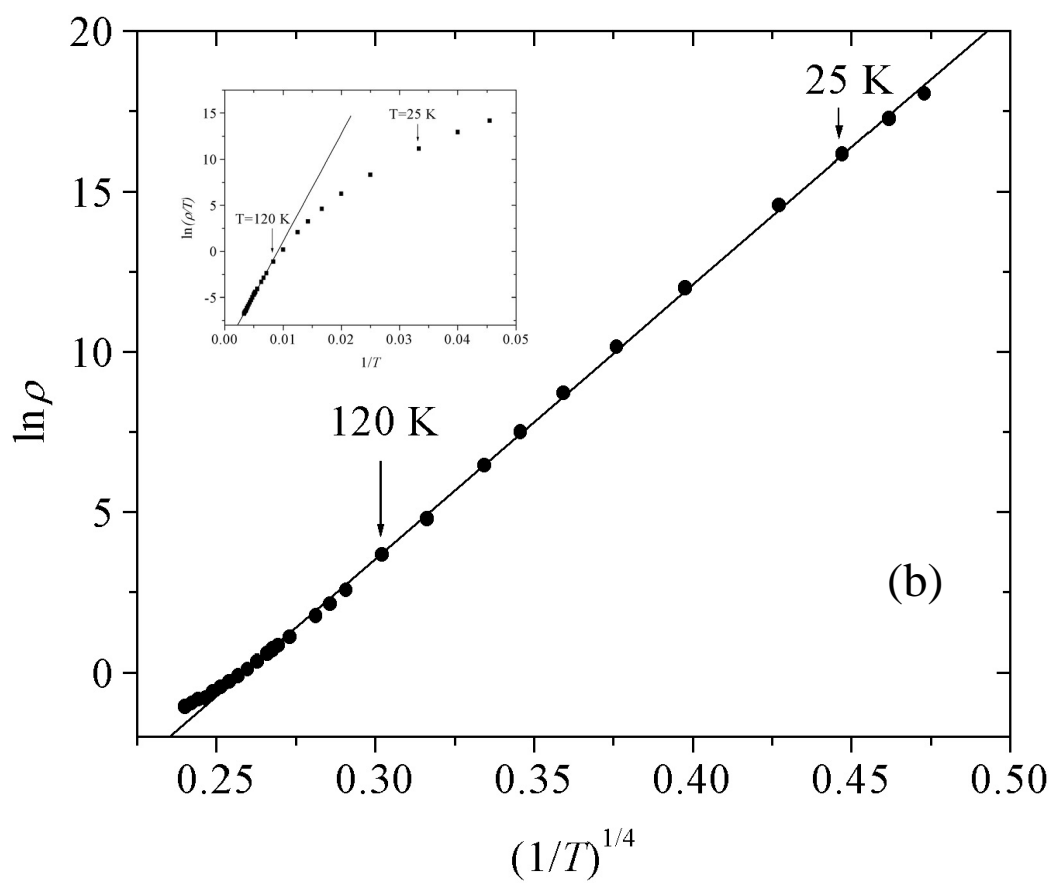

A. Fig. 8

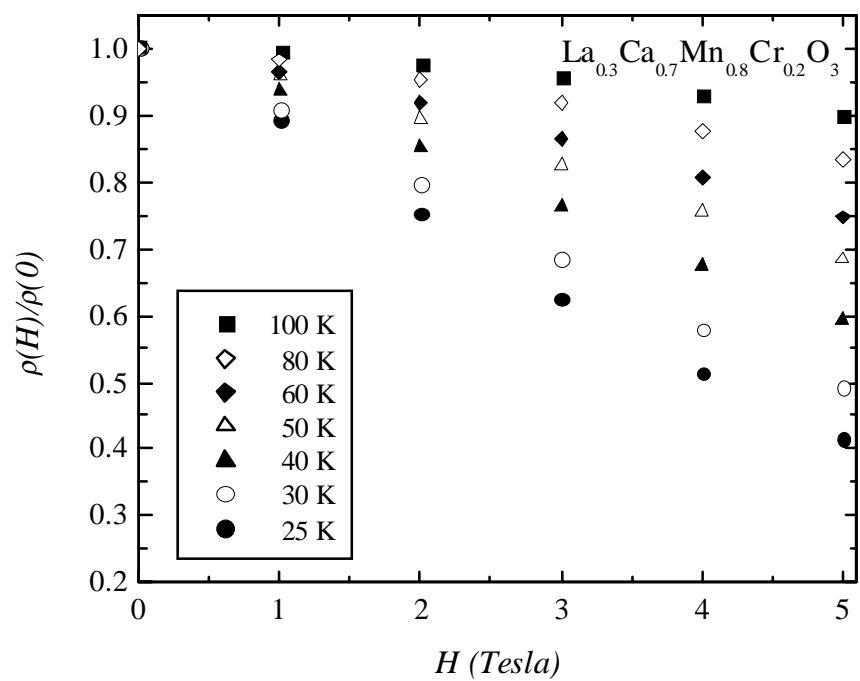

**Fig. 9**

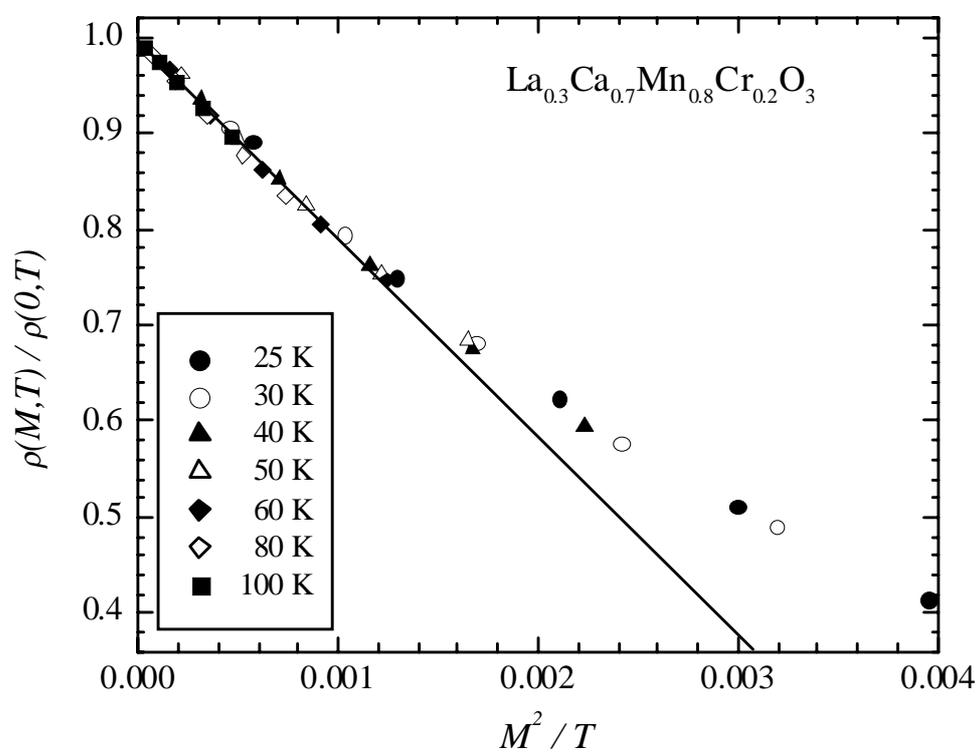

**Fig. 10**